\documentclass[aps,pre,showpacs,preprint]{revtex4-1}

\usepackage{epsfig}
\usepackage{graphics}
\usepackage{graphicx}
\usepackage{epstopdf}
\usepackage{mathrsfs}
\DeclareGraphicsExtensions{,.pdf,.png,.jpg,.gif}

\begin{document}

\title{Inexistence of equilibrium states at absolute negative temperatures}

\author{V\'{\i}ctor Romero-Roch\'in} 

\email{romero@fisica.unam.mx}

\affiliation{Instituto de F\'{\i}sica, Universidad
Nacional Aut\'onoma de M\'exico. \\ Apartado Postal 20-364, 01000 M\'exico D.
F. Mexico. }

\date{\today}

\begin{abstract}
We show that states of macroscopic systems with purported absolute negative temperatures are not stable under small, yet arbitrary, perturbations. We prove the previous statement using the fact that, in equilibrium, the entropy takes its maximum value. We discuss that, while Ramsey theoretical reformulation of the Second Law for systems with negative temperatures is logically correct, it must be a priori assumed that those states are in thermodynamic equilibrium. Since we argue that those states cannot occur, reversible processes are impossible and, thus, Ramsey identification of absolute negative temperatures is untenable.
\end{abstract}


\maketitle

\section{Introduction}

While the Laws of Thermodynamics have shown to be of universal applicability, it is necessary to postulate the existence of equilibrium states in order to develop the theory of thermodynamics in terms of concepts such as temperature and entropy \cite{Fermi}. Without equilibrium states, reversible processes are not possible and, therefore, there is no way to identify the reversible transfer of heat as
\begin{equation}
\delta Q = T dS .
\end{equation}
Ramsey \cite{Ramsey}, without proof, makes this assumption in reformulating the Second Law for systems with energy spectra bounded from above. Then, by logical consistency, negative temperatures appear as a reality. The flaw in such a reformulation is that states with energy above the corresponding one with infinite positive temperature are not stable. The purpose of this article is to show such a result. \\

The Laws of Thermodynamics are, First, the conservation of energy, Third, the unattainability of the Absolute Zero, and Second, the impossibility of, energy conserving, perpetuum mobiles. With these laws and the additional observation that bodies externally unperturbed reach a state of stable equilibrium, one can postulate the existence of reversible processes and define entropy, temperature, chemical potential and, in the case of fluids, hydrostatic pressure; in the case of magnetic systems one can then define the magnetic field (this is implicit in  Maxwell electrodynamics of continuous media \cite{LandauEM}). Let us review the formulation of the Second Law.\\

The Second Law is based on two seemingly different statements, the Clausius and the Kelvin Postulates, which in turn are based in denying the opposite of two empirical observations that we believe are always true. This denial renders impossible a perpetuum  mobile.\\

Clausius Postulate is based on the observation that when two bodies are in contact, energy in the form of heat  spontaneously flows from one to the other or nothing happens. The body that releases the energy it is called {\it hot}, the other {\it cold}. Thus, we say, heat always flows irreversibly from the hotter bodies to the colder ones. If heat does not flow, we say the bodies are in thermal equilibrium. Clausius Postulate is the denial of the opposite, in the form of:  it is impossible to realize a process whose {\it only} result is the transfer of heat from a cold body to a hotter one. \\

On the other hand, Kelvin Postulate is based on the empirical observation that it is always possible to irreversibly convert work into heat, independently of the state of the body. Dissipative friction is the essence of this form of energy transfer. Kelvin Postulate denies the contrary with the statement: it is impossible to realize a process whose {\it only} result is the conversion of heat into work, while  keeping the same state of the body releasing the energy.\\

As described in any textbook, see Ref. \cite{Fermi} for instance, the previous statements are equivalent. They guarantee the impossibility of perpetual motion of energy conserving engines. The validity of the Second Law is based on the fact that we have found no process that violates it. We know from atomic considerations, however, that its validity is only of statistical nature.\\

With the previous Laws {\it and} the hypothesis of the existence of states of thermodynamic equilibrium, one can conceive a {\it reversible} process which is one that consists of equilibrium states only. Then, a hypothetical engine that goes through a cyclic process, a Carnot engine, allows us to prove \cite{Fermi}  a series of statements that lead to the introduction of (1) an Absolute Temperature, which in turn requires the use of the Ideal Gas Temperature, a positive quantity by definition, and (2) Clausius Inequality. This inequality leads to the identification of the Entropy. We insist that these two concepts, temperature $T$ and entropy $S$, require the existence of the state of equilibrium. The theorems needed to prove these assertions are given in the Appendix. \\

An important step in the previous procedure of identifying the temperature is the proof that for {\it all} Carnot engines working between the same two heat sources, the ratio of the extracted $Q_{in}$ to the released $Q_{out}$ amounts of heat are all the same; see the Appendix. Thus, such a ratio must be a property of the heat sources and not of the engines. By appealing to an engine made out of an ideal gas, one reaches the conclusion that,
\begin{equation}
\frac{|Q_{out}|}{|Q_{in}|} = \frac{T_C}{T_H} \label{TS}
\end{equation}
where $T_C$ and $T_H$ are the respective temperatures of the sources with which the body interchanged $Q_{out}$ and $Q_{in}$. The signs of the latter are determined by the direction of the cycle. The subscripts $C$ and $H$ stand for ``cold" and ``hot". It is very important to insist that the variable $T$ is unambiguously determined by an empirical measurement using a (real) gas thermometer. It can also be shown that if $T_C >0$, then, all temperatures must be positive as well. The Third Law ensures that temperatures equal or below  Absolute Zero are unattainable. It is a further simple exercise to show that all reversible Carnot engines working between the same heat sources have the same largest efficiency (less than 1) and that all the corresponding irreversible ones have a lower efficiency.  \\

By a logical argument, that can be found in Ref. \cite{Callen} for instance, one can further prove that the entropy of a thermally isolated system cannot decrease in any process and this in turn implies that the entropy is a {\it concave} function of its arguments, in particular, as a function of the internal energy $E$ of the system. This latter property is typically referred in textbooks as the mathematical statement and requirement that the equilibrium state must be a stable one. That is, it must be stable under small {\it arbitrary} perturbations. We shall return to this point below. It is important to realize that the concave property of the entropy by itself does not indicate that the entropy should be a monotonic increasing function of the energy, thus ensuring positive temperatures only: Twentieth Century Quantum Physics had a surprise for us. We recall here that the temperature is given by,
\begin{equation}
\frac{1}{T} = \left(\frac{\partial S}{\partial E}\right)_{N,X} \label{T}
\end{equation}
where $N$ is the number of atoms (or molecules) and $X$ stands for the values of the extensive appropriate parameters that determine the macroscopic state of the system \cite{LandauSP}. \\

Although Thermodynamics is considered to be a purely empirical science, we now know that we cannot ignore the atomic structure of matter. This was foreseen by Boltzmann \cite{Boltzmann} that found that the Second Law was actually statistical in origin and further found that entropy is given by, 
\begin{equation}
S = k_B \> \ln \Omega(E,N,X) \label{S}
\end{equation}
where $\Omega(E,N,X)$ is the number of ``microscopic" states of a system with $N$ atoms (or molecules) with internal energy $E$ and a given value of the extensive parameters $X$. Thus, $S$ will be monotonous in $E$ if the energy spectrum is unbounded from above. We expect it to be bounded from below by the Ground State Energy, thus incidentally explaining the Third Law. The surprise was that quantum systems with individual entities that have a finite number of states, such as a collection of spins in magnetic materials, give rise to a spectrum for the whole system that is bounded from above. The calculation of the entropy using formula (\ref{S}) gives rise to a function that indeed is concave but that it is not monotonic, since the number of states of the system first raises, reaches a maximum, and then decreases. As there being only one state with the lowest energy, there is also only one with with the highest one. This will be exemplified later on. The consequence of this behavior is that, for values of the energy higher than the one with the largest number of states, the temperature would take a negative value as indicated by the relation (\ref{T}). As it will be further discussed, it is important to stress out that negative temperatures are hotter than all positive ones, because the lay above the positive infinite temperature.\\

\section{Ramsey Postulate of the Second Law}

The previous known result \cite{Purcell} prompted Ramsey \cite{Ramsey} to review the statements of the Second Law such that negative temperatures would not violate it. The main goal of Ramsey was to identify negative temperatures using the relation (\ref{TS}) in a well defined cyclic process. He certainly achieved it, as we show below, but we want to point out a couple of observations right away. First, Ramsey tacitly assumed that equilibrium state always exist and that they are stable. And second, he never specified a ``universal" thermometer that would permit an unambiguous measurement of negative temperatures.  Since Ramsey publication \cite{Ramsey} there have been many articles \cite{Coleman,Landsberg1,Hecht,Geusic,Tykodi1,Machlup,Klein,Tremblay,Danielian,White,Tykodi2,Tykodi3,Landsberg2}, as well as brief sections in well known textbooks \cite{LandauSP,Pippard,Pathria,Huang}, addressing the subject. However, the stability or mere existence of those states has not been discussed and that is the purpose and motivation of this article.\\

In order to use equation (\ref{TS}) to identify negative temperatures, Ramsey modified Kelvin statement. The idea was to reverse the empirical observation and its negation. That is, Ramsey established, first, that heat can always irreversibly be transformed into work irrespective of the state of the body, and then, denied its opposite in the form of: it is impossible to realize a process whose {\it only} result is the transformation of work into heat. This statement certainly forbids an energy conserving perpetuum mobile. In addition, it does allow to construct a cyclic engine that leads to relation (\ref{TS}) with negative temperatures. Furthermore, it can be shown that heat flows from a body at negative temperature to a body at positive temperature, thus proving that the former are hotter than the latter. In the Appendix, we provide the theorems needed to corroborate these statements. But before arguing that states at presumably negative temperature are not stable equilibrium states, we would like to point out that a world with negative temperatures leads to very counterintuitive results.\\

First, we observe that the statement that heat can irreversibly be all converted into work implies that engines operating between reservoirs at different negative temperatures are irrelevant since one can obtain work from any single body, with efficiency equal to one. We could exhaust all the energy of a large body and obtain quite freely a lot of usable work, the dream of clean energy made come true. Additionally, one can prove that all {\it reversible} engines working between negative temperature reservoirs have the {\it lowest} efficiency, but alas, those engines are useless. Secondly, the fact that Ramsey postulates that work cannot be converted into heat, implies that bodies mechanically interacting with others at negative temperatures cannot dissipate energy by friction! thus, they could move through such media as if the body at negative temperature  were a perfect superfluid. Very strange. \\

\section{Instability of states at negative temperatures}

Despite that no violation of the Second Law is allowed, if Ramsey's extension is valid, we can show that systems at negative temperature are not truly stable. That is, we can show that if a system in a state of negative temperature interacts with another one that can only take positive temperatures, the equilibrium state always has a positive temperature, {\it independently} of the sizes of the systems. Consider the extreme case that the ``normal" body is much smaller than the one at negative temperature, such that it should be considered merely as a perturbation to the latter. As we see, however, it is far from being a perturbation, since it will always drive the large system out of the state of negative temperature and will equilibrate it into a state with a positive temperature. Thus, states with negative temperatures are not stable under small, otherwise arbitrary perturbations. This result follows quite simply from the Second Law due to the unboundedness of the number of states of systems that have can only positive temperatures.\\

To be more precise, let us call $p$ the system that can only take positive temperatures and $n$ the one that can take both. Suppose the latter is in a macroscopic state that should have a negative temperature. Let $E_n$, $N_n$ and $X_n$ be the energy, number of particles and appropriate extensive variables of the $n$ system. Let also $S_n(E_n,N_n,X_n)$ be the entropy of the state of $n$, as given formally by equation (\ref{S}). Analogously for $p$, we identify $S_p(E_p,N_p,X_p)$. Consider the situation that the systems are in thermal contact but isolated from any other body. In this case, the total energy is a constant, $E_T = E_p + E_n$, but the systems interchange it until the state of equilibrium is reached. This state corresponds to that one with the maximum value of the total entropy $S_T = S_p + S_n$. Since the interaction is only through thermal contact all the other variables but the energy remain constant. Let $E_p^{eq}$ and $E_n^{eq}$ be the energies at the common equilibrium state. They obey, of course, $E_T = E_p^{eq} + E_n^{eq}$. It is a straightforward exercise to show that in the state of maximum entropy it is true that
\begin{equation}
\left(\frac{\partial S_p}{\partial E_p^{eq}} \right)_{N_p,X_p}= \left(\frac{\partial S_n}{\partial E_n^{eq}} \right)_{N_n,X_n} \label{equi}
\end{equation}
which, by equation (\ref{T}), simply says that the temperatures of the two bodies are the same. Since the system $p$ can only take positive temperatures, the common one must be positive. The argument is independent of the sizes of the systems.\\

What is not independent of the sizes of the systems is the heat capacity, whose behavior is tied to the stability of the state of equilibrium. It is almost common sense to affirm that when two bodies at different temperatures are put into thermal contact, the equilibrium temperature not only is between the initial ones but that it is closer to the body with the largest heat capacity. Since the heat capacity scales with the size of the system, the final temperature is closer to the temperature that the bigger body had initially. As a matter of fact, a heat reservoir is nothing but a very large body such that its heat capacity is also so large that it keeps essentially its same temperature regardless of the heat released to, or absorbed from, a smaller system. As we have seen in the previous paragraph, a body at negative temperature does not behave in this ``common" way. Moreover, the fact that the heat capacity must be positive as a consequence of the stability of the equilibrium state, is most clearly summarized by the Second Law in the form of Le Chatelier's principle which establishes that: ``an external interaction which disturbs the equilibrium brings about processes in the body which tend to reduce the effects of this interaction" \cite{QLL}. For a large body at a negative temperature in interaction with a smaller system that can take positive ones only, one finds what appears to be the opposite, namely, there must be internal processes that tend to take the body out of that equilibrium state. It is in this regard that we can assert that the states of seemingly negative temperature cannot be in thermal equilibrium.\\

Let us visualize the previous general results with a very simple example. Let $n$ be an ideal paramagnetic solid of spin $j = 1/2$ in the presence of a uniform magnetic field, whose Hamiltonian can be written as,
\begin{equation}
H_n = - \mu_0 B \sum_{i=1}^{N_n} \>m_i \label{H}
\end{equation}
where $\mu_0$ is the magnetic moment of the atoms, $B$ the external magnetic field and $m_i = \pm 1/2$ the spin component along $B$. For the $p$ system we take a monoatomic ideal gas of fermions of mass $m$ and spin $j=1/2$. In both cases it is straightforward  to calculate \cite{details} the entropies $S_p(E_p,N_p,V_p)$ and $S_n(E_n,N_n)$. For ease of plotting we use arbitrary energy units with $\mu_0 B = 15$. We consider 10 moles of the paramagnet and 1 mole of the ideal gas, namely,  $N_n = 10 N_0$ and $N_p = N_0$, with $N_0$ Avogadro number, in order to ensure the condition $N_n \gg N_p$.\\

\begin{figure}
\begin{center}
\includegraphics[width=0.6\textwidth]{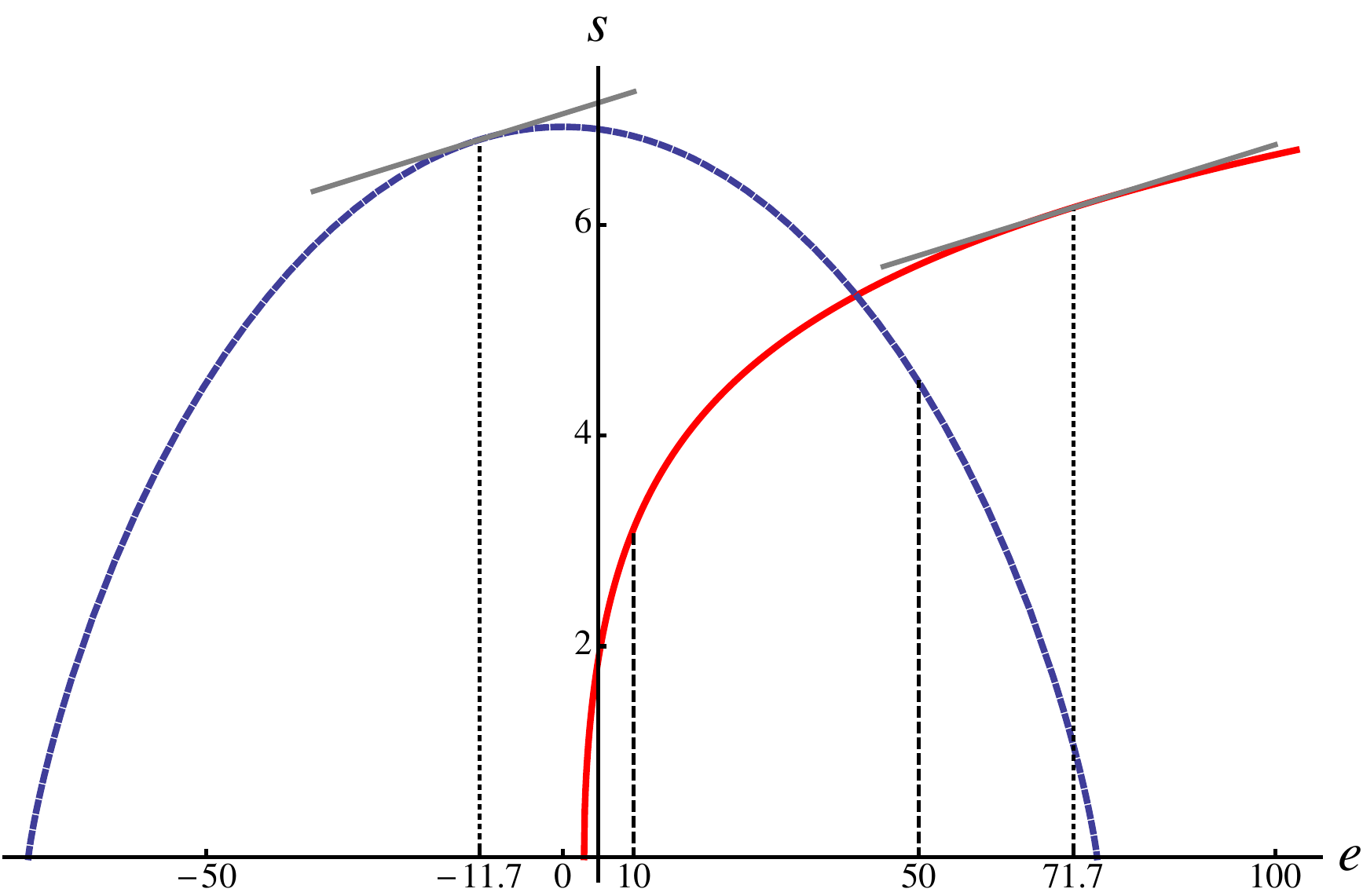}
\end{center}
\caption{(Color online) Entropies $s = S/N_0 k_b$ versus energy $e = E/N_0$, for a paramagnet $n$ of ten moles $N_n = 10 N_0$, thick-dashed blue line, and for one mole $N_p = N_0$ of an ideal Fermi gas $p$, solid red line. The energy units are arbitrary with $\mu_0 B = 15$. The volume of the gas was chosen such that Fermi energy is $\epsilon_F = 5$. The dashed vertical lines indicate the initial energies $e_n = 50$ and $e_p = 10$. The equilibrium state is indicated with vertical dotted lines and correspond to $e_n^{eq} = -11.7$ and $e_p^{eq}=71.7$. At those states we have plotted the slope of the curves $s$ vs $e$ respectively, to show that correspond to the same temperature $k_B T = 47.6$.}
\label{F1}
\end{figure}

\begin{figure}
\begin{center}
\includegraphics[width=0.6\textwidth]{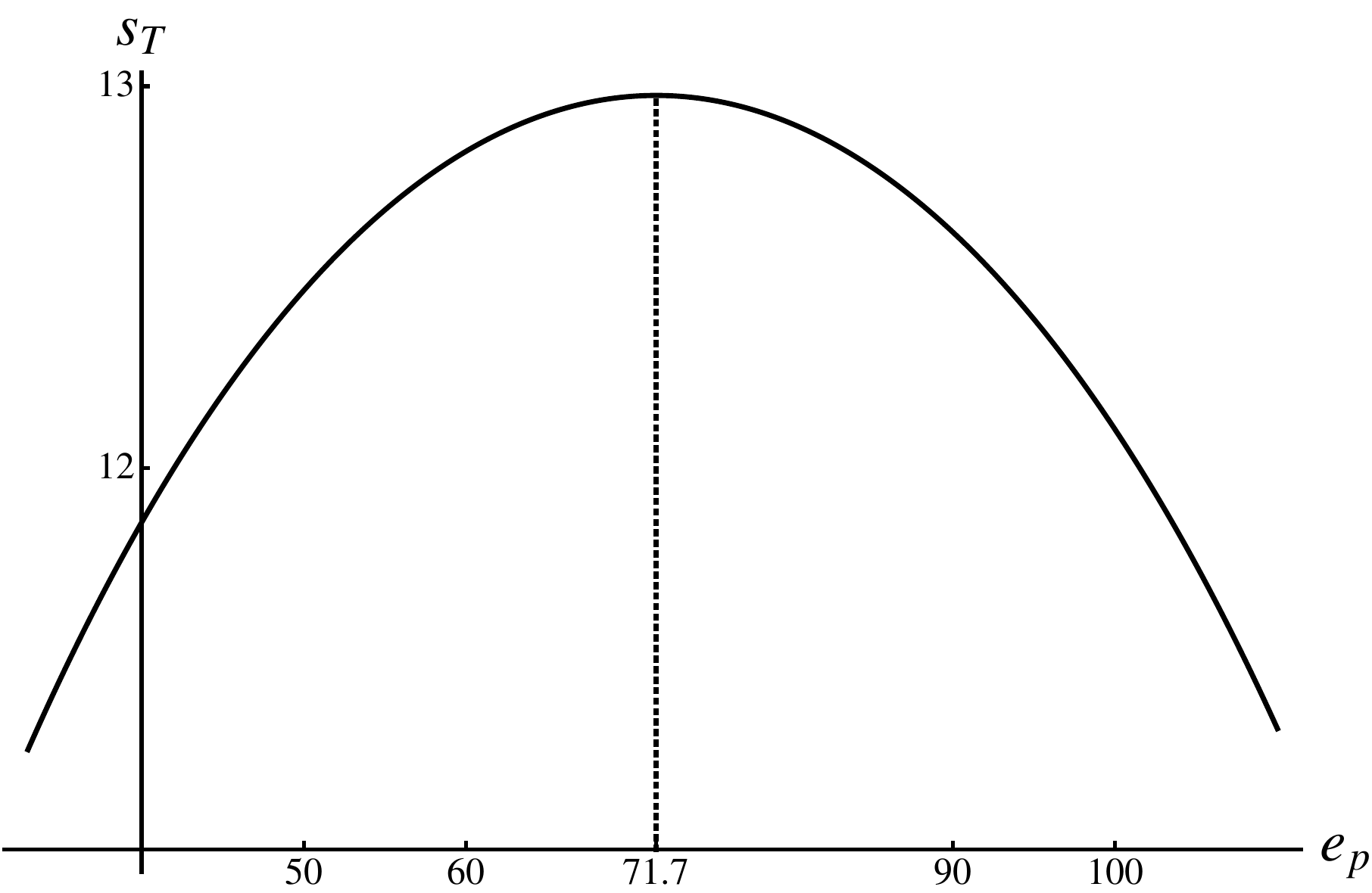}
\end{center}
\caption{Total entropy $s_T = S_T/N_0 k_B$ versus energy $e_p = E_p/N_0$ of the system $p$ (ideal gas). The maximum occurs at $e_p^{eq} = 71.7$, see Figure \ref{F1}.}
\label{F2}
\end{figure}

The physical situation is that, before thermal contact, the $p$ system is at a ``low''  positive temperature with energy $E_p/N_0 = 10$, while the paramagnet is at a negative temperature with energy $E_n/N_0 = 50$. In Figure \ref{F1} we plot the entropies $S_n/N_0 k_B$ (blue) and $S_p/N_0 k_B$ (red) versus energy $E/N_0$. The initial energies are indicated with dashed lines. Further numerical details are given in the caption. When the systems are put into thermal contact they equilibrate at the same temperature, as indicated by the relation (\ref{equi}). At the equilibrium state the systems have the energies $E_n^{eq}/N_0 = -11.7$ and $E_p^{eq}/N_0 = 71.7$; these are indicated with dotted lines. The common equilibrium temperature is $k_B T = 47.6$, a large, positive one. This is illustrated by showing that at the equilibrium state the slopes are the same for the $n$ and $p$ systems.  It is instructive to verify that the total entropy takes a maximum at those values of the energies. This is shown in Figure \ref{F2} where we plot the total entropy as a function of the energy $E_p/N_0$; the energy of the paramagnet is not an independent variable since $E_n = E_T - E_p$. In this example,  $E_T/N_0 = 60$. \\

This simple geometric construction shows that the total entropy, for fixed total energy, of a system that can only take positive temperatures in interaction with a system that can take both positive and negative temperatures will always  take its maximum value at a state with a common positive temperature. The reason is that, since the number of states is not bounded for a $p$ system, it will always be possible to increment the total number of states using this system. The system $n$ is ``oblivious" to this change because its number of states is bounded from above. We repeat, it could be that $N_p \ll  N_n$, as in the figures, and then it will always happen that the system at negative temperature (which is very hot!) will cool down to achieve a positive temperature and the small system will certainly get hot, but at that positive temperature.\\

The conclusion is, therefore, that a system in a macroscopic state to which it would correspond a negative temperature cannot relax to, neither stay at, a stable equilibrium state. It will always transfer its energy to any ``normal" body with which interacts and necessarily leave that state. It is only in this respect that one can consider those states as being ``hotter" than any body at positive temperatures, but it does not mean at all that we can assign a negative temperature. A very important point here is that we are not simply implying that it is experimentally ``very difficult" to maintain a system in a state of negative temperature. We are asserting that it is a matter of principle. That is, the concept of ``thermal isolation" is meaningful only for systems at positive temperatures: With extreme experimental care, it is only possible for those states to reach a situation for which the interaction of the system with the rest of the Universe amounts to a very small perturbation. This is impossible for systems at negative temperature.
In the absence of those equilibrium states, the existence of reversible processes cannot be postulated. And as discussed above, this precludes the definition of temperature and entropy. Although not argued here, one runs into similar conclusions when considering the internal mechanisms of thermal relaxation to the equilibrium state.\\

This discussion of course does not imply that systems cannot be put into macroscopic states that would correspond to negative temperatures. As a matter of fact, it has been done in magnetic systems \cite{Purcell} and, very recently, in ultracold gases in optical lattices \cite{Bloch}. However, as described above, those states are necessarily unstable.  Nevertheless, to reinforce this point, it is also of interest to discuss possible experimental realizations to highlight situations that appear contradictory or even absurd. We shall argue that those apparent difficulties are solved by realizing that one made an invalid hypothesis about equilibrium states at the outset.\\

\section{Reaching states with $T < 0$ by heating or cooling}

To argue about the title of this Section, let us assume that states at negative temperatures would indeed be of stable equilibrium. As pointed out in all articles regarding negative temperatures, these are hotter than positive ones, and thus one can reach them by ``heating" the system, or equivalently, by putting energy into it. But, is this always true? One should recall here that in the mid-70's there was a flurry of papers\cite{Tykodi1,Tremblay,Danielian,White,Tykodi2} discussing the impossibility not only of {\it reversibly} reaching the ``usual" coldest temperature $T = +0$, but also of reversibly attaining $T =\pm \infty$ and $T = -0$, the hottest possible temperature. For the ideal paramagnet of Eq.(\ref{H}) these results are reflected in the fact that the heat capacity at constant  $B$, $C_B = N_n k_B (\mu_0 B/k_B T)^2 {\rm sech}^2 (\mu_0 B/2k_B T)$, vanishes both at $ T = \pm 0$ and at $T = \pm \infty$. However, as the original experiment by Purcell and Pound \cite{Purcell} showed, it is possible to pass from positive to negative temperatures, or viceversa, by an {\it irreversible} process. Let us use the simple ideal paramagnetic system described before, Eq. (\ref{H}), to discuss two possible processes to show that one can ``reach" negative temperatures from positive ones by either heating, as expected, or by cooling! The ``trick", we shall see, is that the  states $T = +0$, $T = \pm \infty$ and $T = -0$ are all unattainable and, thus, it appears meaningless to consider ``heating" or ``cooling" above or below them. Those states must always be ``bypassed". By the way, this forbids the reversible operation of an engine working between reservoirs at temperatures with different signs. \\

\begin{figure}
\begin{center}
\includegraphics[width=0.8\textwidth]{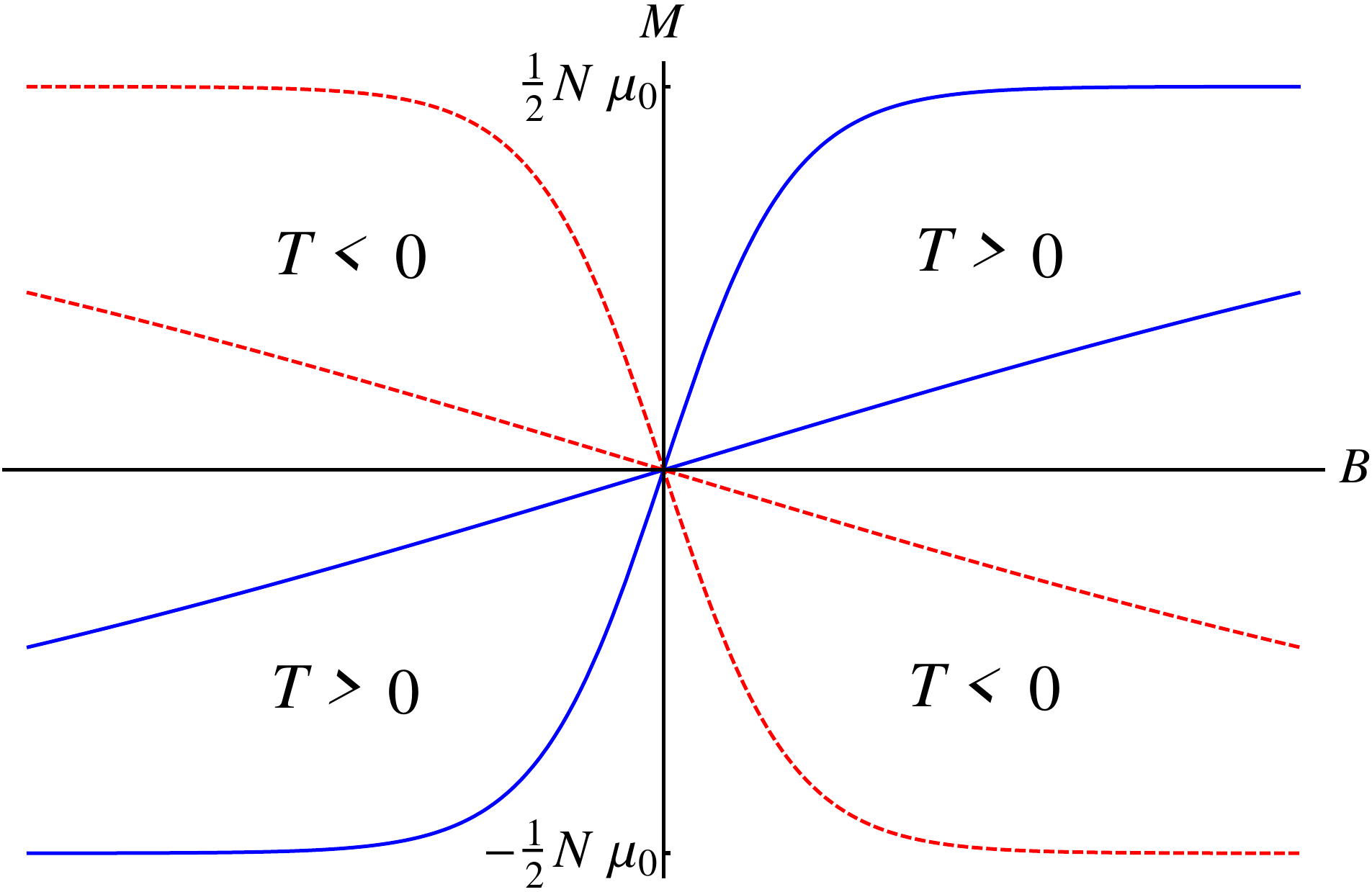}
\end{center}
\caption{(Color online) Equation of state $M = M(B,T)$ of an ideal paramagnet, see Eq.(\ref{M}). Positive temperatures correspond to the magnetization $M$ and magnetic field $B$ being parallel (solid blue curves), while negative ones to the antiparallel situation (dashed red curves).}
\label{EOS}
\end{figure}

\begin{figure}
\begin{center}
\includegraphics[width=0.8\textwidth]{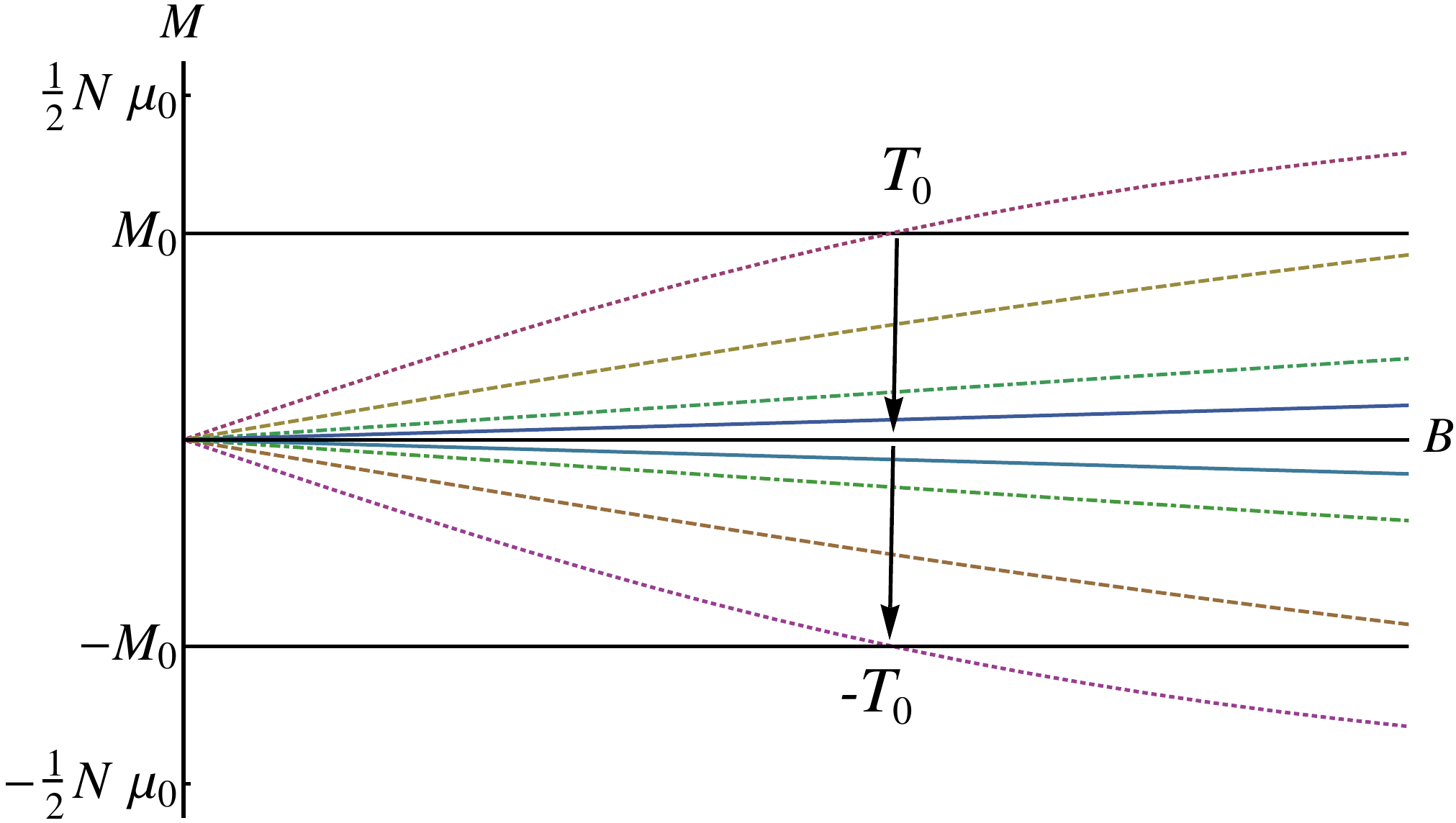}
\end{center}
\caption{(Color online) Change of magnetization $M$ for a fixed value of the magnetic field $B$. Initially, the system has magnetization $M_0$ at positive temperature $T_0$. The system is reversibly heated up to $T \to \infty$, indicated by the upper arrow. Then, irreversibly, the magnetization is inverted reaching a temperature $T \to -\infty$. Afterwards, it continues to be reversibly heated up to $-T_0$ and $-M_0$.}
\label{Bcte}
\end{figure}

\begin{figure}
\begin{center}
\includegraphics[width=0.8\textwidth]{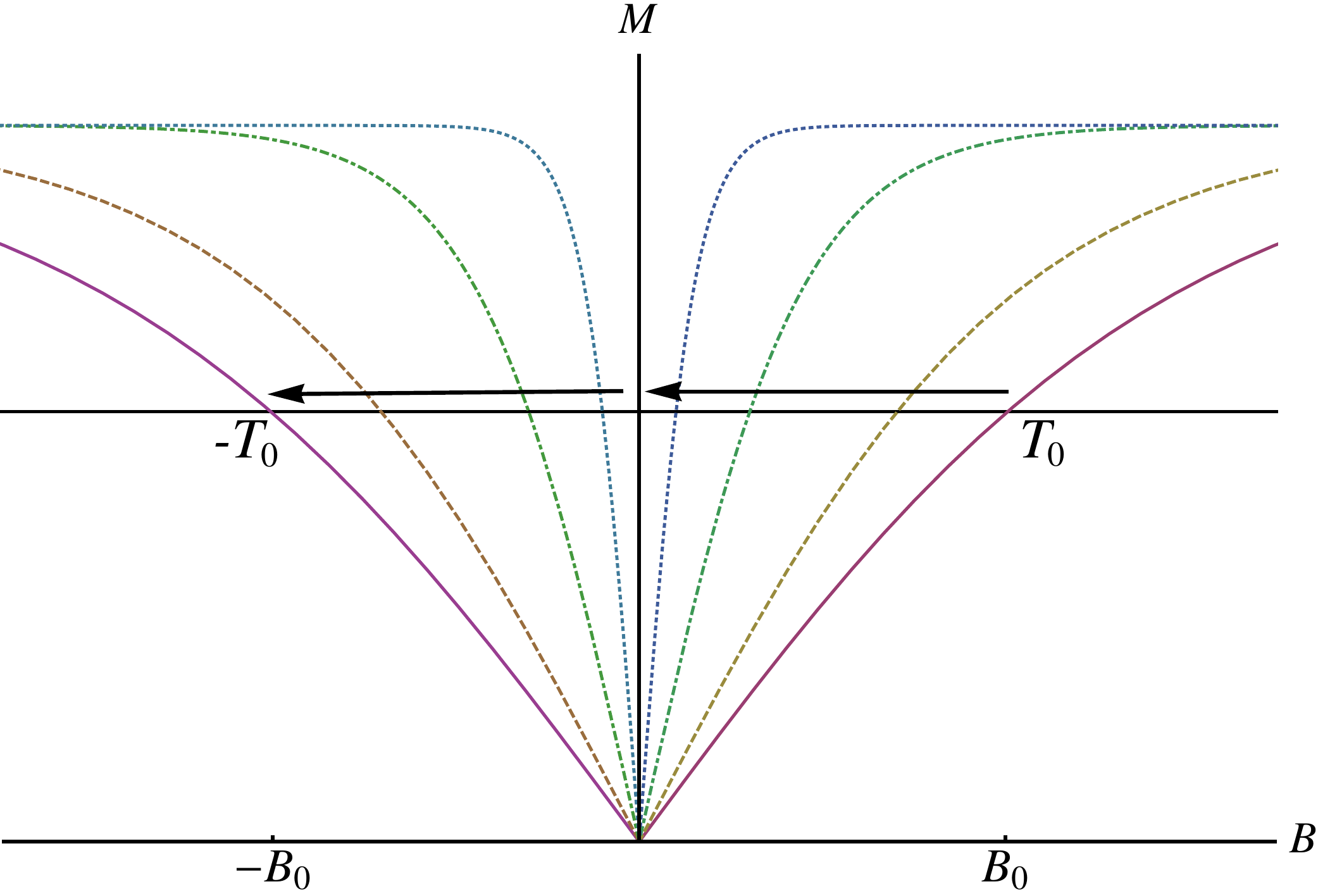}
\end{center}
\caption{(Color online) Change of magnetic field $B$ for a fixed value of the magnetization $M$. Initially, the system has magnetic field $B_0$ at positive temperature $T_0$. The system is reversibly cooled down to $T \to +0$ at constant magnetization, as indicated by the right arrow. Then,  the field is irreversibly inverted reaching a temperature $T \to -0$.  After that, it continues to be reversibly cooled down to $-T_0$ and $-B_0$.}
\label{Mcte}
\end{figure}

Considering the ideal paramagnet described by Eq. (\ref{H}), one can find the equation of state of the system that relates the magnetization $M$ to the magnetic field $B$ and temperature $T$, namely $M = M(B,T)$:
\begin{equation}
M = \frac{1}{2} N \mu_0 \> \tanh \left( \frac{\mu_0 B}{2 k_B T}\right) .\label{M}
\end{equation}
This formula is valid for any sign of the field $B$ and the temperature $T$. This equation of state is shown in Fig. \ref{EOS}, where we see that positive temperatures correspond to states with the magnetization $M$ being parallel to the magnetic field $B$, while at negative  temperatures those variables are antiparallel. Let us now think of two simple processes starting at positive temperature and reaching negative ones by, (I) change of magnetization $M$ at constant field $B_0$, as shown in Fig. \ref{Bcte}, and (II) change of field $B$ at constant $M_0$, Fig. \ref{Mcte}. \\

In the first case, Fig. \ref{Bcte}, the system can be reversibly heated as much as one desires, reaching an extremely large positive temperature, but not quite $T = + \infty$, with an almost vanishing magnetization. Then, as Purcell and Pound did, the magnetization can be irreversibly inverted, bypassing $T = \pm \infty$, and hence obtaining a very large negative temperature. The system can then be further reversibly heated up with negative temperature reservoirs. This scenario appears sound in the sense that negative temperatures are reached by {\it heating} the system. The realization of negative temperature reservoirs is, at this point, considered feasible.\\

The second case, Fig. \ref{Mcte}, refers to a reversible decrease of the magnetic field $B$ at constant magnetization. Incidentally, for this simple system this process is also adiabatic, with a corresponding positive change in energy. The system can then be cooled down reversibly to reach almost $T = +0$, as cold as we want. At those very low positive temperatures the magnetic field is also extremely small. Again, we cannot reversibly reach $T = +0$ or below, but nothing prevents us from {\it irreversibly} ``cooling" even further by inverting ``very quickly" the direction of the very feeble magnetic field $B$ and putting the system at a negative temperature very close to $T = -0$ ... but this is the hottest temperature in the Universe! So, was the magnet cooled down or heated up? if the latter, the system jumped from the absolute coldest temperature to the absolute hottest one by simply inverting a vanishing field $B$. We note that the change in energy is also very tiny but positive, $\Delta E = 2 M_0 B$, indicating a cooling mechanism. This process does not contradict the fact that negative temperatures are hotter than positive ones, but it indicates that the transition from positive to negative temperatures can be achieved by an irreversible cooling mechanism {\it below} Absolute Zero.\\

The two previous processes indicate that, although negative temperatures are hotter than positive ones in the sense of the direction of heat flux, one can pass from one case to the other through $T = + \infty$ or $T = + 0$. As mentioned above, there appears to be no formal difficulty since the transition must be irreversible. But from a more practical point of view, in which reversible processes actually never occur, it appears absurd and contradictory that negative temperatures can be found above $T = +\infty$ or below $T =+0$, depending on the process used ... It seems that, somewhere, something is wrong. However, if we accept that the purported negative-temperature states are not equilibrium states (and no negative reservoirs can thus exist) then a solution appears. As a matter of fact, it is very simple to conclude that in reality, in the ``cooling" process, the paramagnet will eventually equilibrate with the negative-pointing magnetic field, but at a very low {\it positive} temperature. It will do so by any minute interaction with the rest of the Universe. The possibility of irreversibly reaching those states is not denied, but it appears that the concepts of temperature and entropy applied to them, with all their concomitant properties, are meaningless. It remains true, nevertheless, that those non-equilibrium states are certainly ``hot" because they will eventually release their energy. But at the end, they will reach an equilibrium state with a ``cold" positive-temperature.\\ 

\section{Final Remark}

An interesting question arises if one imagines a system that can take negative temperatures {\it only}. This is claimed to have been realized in the recent experiment with ultracold gases in optical lattices \cite{Bloch}, based on previous theoretical suggestions \cite{Mosk,Rapp}. In those articles it is described how, by carefully  inverting the sign of atomic and external interactions, the atomic kinetic energy also becomes effectively negative. The ensuing system has an energy spectrum with an upper bound, but unbounded from below.  Incidentally, the purported negative-temperature state was obtained by cooling the gas below $T = +0$, similarly to the process of Fig. \ref{Mcte}. Then, what would happen if such a system were put in contact with a system that could only take positive temperatures? (as essentially the rest of the Universe). Since energy always flows from the hotter to the colder, according to Clausius, then the possible scenario is that energy will continuously be transferred from the system at negative temperature to the one at positive ones. That is, the former will cool down towards $T \to -\infty$, while the latter will heat up until $T \to +\infty$, but truly never equilibrating. This kind of behavior again appears quite strange and does not seem to be observed in the recent experiments. We insist that a system in such a situation requires negative kinetic energy, a concept whose fully elucidation and implications demands further thought.

\appendix*

\section{Thermodynamic theorems.}

In this Appendix, we summarize the basic thermodynamic theorems that lead to absolute positive or negative temperatures, if either Kelvin or Ramsey postulates are used. The proofs are not given. They can be produced by following the arguments given by Fermi \cite{Fermi}. \\

We shall denote by $H$ a hot reservoir and by $C$ a cold one. The Second Law is given by:\\

\noindent
$\bullet$ {\bf Clausius postulate} (CP). While heat can be irreversible transferred from $H$ to $C$, it is impossible  the opposite as a sole result. 

\noindent
$\bullet$
{\bf Kelvin postulate} (KP). While work can always be converted into heat, $W \to Q$, it is impossible  the opposite as a sole result, $Q \not\rightarrow W$.

\noindent
$\bullet$
{\bf Ramsey postulate} (RP). While heat can always be converted into work, $Q \to W$, it is impossible  the opposite as a sole result, $W \not\rightarrow Q$. \\

\noindent
{\bf Theorem 0.} CP is equivalent to KP or CP is equivalent to RP. Note that RP is not the negation of KP, but rather its inverse. Thus, NO KP is the same as NO RP and, therefore, both are equivalent to NO CP. It is clear, however, that either KP or RP cannot be true simultaneously.\\

\noindent
{\bf Carnot engine.} A Carnot engine is a device that operates in a cycle between an $H$ and a $C$ reservoirs. Let $Q_H$ and $Q_C$ be the corresponding amounts of heat interchanged between reservoirs and the engine, and $W$ the work received or delivered by the engine, all in one cycle. By the First Law, it is true that, 
\begin{equation}
W + Q_H + Q_C = 0. \label{FirstL}
\end{equation}
The signs of $W$ and $Q$'s are negative if released by the engine. Note that a Carnot engine may or may not be reversible. Namely, neither case violates the Second Law. Reversible means that if operated in any direction, the signs of $W$, $Q_H$ and $Q_C$ are reversed, but their magnitudes remain the same. The efficiency $\eta$ of a Carnot engine is the magnitude of the ratio of the work delivered to the heat absorbed, both by the engine, in one cycle. By the First Law, Eq. (\ref{FirstL}), the efficiency is less or equal than 1.\\

\noindent
{\bf Theorem 1K.} If KP is valid, then for a Carnot engine for which $W < 0$, it must be true that $Q_H > 0$ and $Q_C < 0$. That is, if a Carnot engine delivers work $|W|$, then it absorbs heat $Q_H$ from the hot reservoir and releases heat $|Q_C|$ to the cold one. {\bf Corollary.} The efficiency of this engine is necessarily less than 1, whether reversible or not.\\

\noindent
{\bf Theorem 1R.}  If RP is valid, then for a Carnot engine for which $W > 0$, it must be true that $Q_H > 0$ and $Q_C < 0$. That is, if a Carnot engine receives work $|W|$, then it absorbs heat $Q_H$ from the hot reservoir and delivers heat $|Q_C|$ to the cold one. {\bf Corollaries.} The opposite engine thus delivers work while absorbing heat from the cold reservoir and releasing part of it to the hot one. With the use of CP, the cold reservoir $C$ can be restored to its initial state by transferring heat from $H$. Thus, one can convert all the heat into work, in agreement with RP. This last process is of efficiency 1, yet irreversible.\\

\noindent
{\bf Theorem 2K.} Let KP be valid. Let $W < 0$, $Q_H > 0$ and $Q_C < 0$ refer to a reversible Carnot engine and let $W^\prime < 0$, $Q_H^\prime  > 0$ and $Q_C^\prime < 0$ refer to a Carnot engine not necessarily reversible. Both engines operate between the same $H$ and $C$ reservoirs. Then, it is true that,
\begin{equation}
\frac{|Q_C|}{Q_H} \le \frac{|Q_C^\prime|}{Q_H^\prime} < 1 ,\label{K2}
\end{equation}
where the equality refers to the case when both engines are reversible. The fact that the ratio of those heats is less than one, follows from the First Law, Eq. (\ref{FirstL}). {\bf Corollaries.} If both engines are reversible, then the ratio $|Q_C|/Q_H$ is a property of the reservoirs and not of the engine. This allows to define the temperature $T$ as a property of the reservoirs. Call them $T_H$ and $T_C$ respectively. Because the ratio $|Q_C|/Q_H = T_C/T_H < 1$, then, if $T_C >0$, all temperatures are positive, $T >0$. By running a Carnot engine made out of an ideal gas, the temperature $T$ can be identified with the ideal gas temperature. All reversible engines have the same efficiency, larger than the efficiency of any irreversible one. By considering an arbitrary cycle, Clausius inequality follows.\\

\noindent
{\bf Theorem 2R.} Let RP be valid. Let $W > 0$, $Q_H > 0$ and $Q_C < 0$ refer to a reversible Carnot engine and let $W^\prime > 0$, $Q_H^\prime  > 0$ and $Q_C^\prime < 0$ refer to a Carnot engine not necessarily reversible. Both engines operate between the same $H$ and $C$ reservoirs. Then, it is true that,
\begin{equation}
\frac{|Q_C|}{Q_H} \ge \frac{|Q_C^\prime|}{Q_H^\prime} > 1 ,\label{K2}
\end{equation}
where the equality refers to the case when both engines are reversible. The fact that the ratio of those heats is greater than one, follows from the First Law, Eq. (\ref{FirstL}). {\bf Corollaries.} If both engines are reversible, then the ratio $|Q_C|/Q_H$ is a property of the reservoirs and not of the engine. This allows to define the temperature $T$ as a property of the reservoirs. Call them $T_H$ and $T_C$ respectively. Because the ratio $|Q_C|/Q_H = T_C/T_H > 1$, then if $T_H < 0$, all temperatures are negative, $T < 0$. All reversible engines have the same efficiency, smaller than the efficiency of any irreversible one. By considering an arbitrary cycle, Clausius inequality follows.\\

\noindent
{\bf Theorem 3.} Heat can flow irreversibly from a reservoir at negative temperature $T_N < 0$ to a reservoir at a positive one $T_P > 0$. {\bf Proof.} By RP let us convert a quantity of heat  $Q$ from the reservoir at $T_N$ into work $W$. Then, by KP all this work can be converted into heat in a reservoir at $T_P$. Both RP and KP prohibit the inverse process. The final result is that we were able to irreversibly transfer, as a sole result, the heat $Q$ from the reservoir at $T_N$ to the reservoir at $T_P$. By CP, the reservoir $T_N$ is hotter than the reservoir $T_P$. 

\begin{acknowledgments}

Support from grant DGAPA UNAM IN108812 is acknowledged. I thank the students of the lectures on {\it Heat, Waves and Fluids} I delivered at the 2012 Fall Semester at the School of Sciences at UNAM for motivating me to write this note.

\end{acknowledgments}


\begin{thebibliography}{99}

\bibitem{Fermi} E. Fermi, {\it Thermodynamics}, Dover (1956).

\bibitem{Ramsey} N. Ramsey, Phys. Rev. {\bf 103}, 20 (1956).

\bibitem{LandauEM} L.D. Landau, E.M. Lifshitz and L.P. Pitaevskii , {\it Electrodynamics of Continuous Media}, Pergamon (1984).

\bibitem{Callen} H.B. Callen, {\it Thermodynamics}, Wiley (1960).

\bibitem{LandauSP} L.D. Landau and E.M Lifshitz,  {\it Statistical Physics I}, Pergamon (1980).

\bibitem{Boltzmann} L. Boltzmann, {\it Lectures on Gas Theory}, Dover (1995).

\bibitem{Purcell} E.M. Purcell and R.V. Pound, Phys. Rev. {\bf 81}, 279 (1951).

\bibitem{Coleman} B.D. Coleman and W. Noll, Phys. Rev. {\bf 115}, 262 (1959).

\bibitem{Landsberg1} P.T. Landsberg, Phys. Rev. {\bf 115}, 518 (1959).

\bibitem{Hecht} C.E. Hetch, Phys. Rev. {\bf 119}, 1443 (1960).

\bibitem{Geusic} J.E. Geusic, E.O. Schulz-DuBois, and H.E.D. Schovil, Phys. Rev. {\bf 156}, 343 (1967).

\bibitem{Tykodi1} R.J. Tykodi, Am. J. Phys. {\bf 43}, 271 (1975).

\bibitem{Machlup} S. Machlup, Am. J. Phys. {\bf 43}, 991 (1975).

\bibitem{Klein} A.G. Klein, Am. J. Phys. {\bf 43}, 1100 (1975).

\bibitem{Tremblay} A.-M. Tremblay, Am. J. Phys. {\bf 44}, 994 (1976).

\bibitem{Danielian} A. Danielian, Am. J. Phys. {\bf 44}, 995 (1976).

\bibitem{White} R.H. White, Am. J. Phys. {\bf 44}, 996 (1976).

\bibitem{Tykodi2} R.J. Tykodi, Am. J. Phys. {\bf 44}, 997 (1976).

\bibitem{Tykodi3} R.J. Tykodi, Am. J. Phys. {\bf 46}, 354 (1978).

\bibitem{Landsberg2} P.T. Landsberg, R.J. Tykodi, A.-M. Tremblay, J. Phys. A: Math. Gen. {\bf 13}, 1063 (1980).

\bibitem{Pippard} A.B. Pippard, {\it The Elements of Classical Thermodynamics}, University Press (1961).

\bibitem{Pathria} R.K. Pathria, {\it Statistical Mechanics}, Butterworth-Heinemann (1996).

\bibitem{Huang} K. Huang, {\it Introduction to Statistical Physics}, Taylor-Francis (2001).

\bibitem{QLL} See Section 22 in Ref. \cite{LandauSP}.

\bibitem{details} The entropy of a paramagnet as given by Eq.(\ref{H}) is,
$$S = \frac{N k_B}{2} \left(\frac{2 E}{N \mu_0 B} -1\right) \ln \frac{1 - \frac{2 E}{N \mu_0 B}}{1+\frac{2 E}{N \mu_0 B}}
+ Nk_B \ln \frac{2}{1+\frac{2 E}{N \mu_0 B}} .$$
The entropy of a gas of fermions is given by
$$S = \frac{5}{3} k_B \beta E - k_B \alpha N$$
were the parameters $\alpha = \beta \mu$ and $\beta = 1/k_B T$ are found from,
$$N = 2 \frac{V}{\lambda^3} f_{3/2}(\alpha) \>\>\>{\rm and}\>\>\> E = 3 k_B T \frac{V}{\lambda^3} f_{5/2}(\alpha)$$
with $\lambda = h / \sqrt{2 \pi m k_B T}$ and the Fermi functions given by,
$$f_\nu( \alpha) = \frac{1}{\Gamma(\nu)} \int_0^\infty \frac{x^{\nu -1} dx}{e^{x-\alpha} + 1}.$$

\bibitem{Bloch} S. Braun, J.P. Ronzheimer, M. Schreiber, S.S. Hodgman, T. Rom, I. Bloch, and U. Schneider, Science {\bf 339}, 52 (2013).

\bibitem{Mosk} A.P. Mosk, Phys. Rev. Lett. {\bf 95}, 040403 (2005).

\bibitem{Rapp} A. Rapp, S. Mandt, and A. Rosch, Phys. Rev. Lett. {\bf 105}, 220405 (2010).

\end{thebibliography}
\end{document}